# On the derivation of reduction from the Schrödinger equation.

# A disproof of no-go theorems and a proposal

Roland Omnès*
Laboratoire de Physique Théorique[#]
Université de Paris XI, Bâtiment 210, F-91405 Orsay Cedex, France

**Abstract**

The possibility of a fundamental consistency between the basic quantum principles and reduction (so-called wave function reduction) is reexamined. The mathematical description of an organized macroscopic device is constructed explicitly as a convenient tool for this investigation. Previous no-go theorems excluding consistency are disproved on this ground, because their assumptions neglected the occurrence of organization in a real measuring apparatus. A scheme for deriving reduction from quantum mechanics is also proposed as a conjecture.

## 1. Introduction and outlook

One reexamines in this work the question of consistency between the basic principles of quantum mechanics and reduction (often called wave function reduction). Consistency would mean then that reduction is a fundamental consequence of the principles, not extraneous to them, as occurred when similar returns to principles removed other difficulties in interpretation. Decoherence, for instance, removed macroscopic interferences [1-6.], consistent histories disposed of most logical paradoxes [7-9], and explicit derivations of emergent classical dynamics explained classical determinism [10] . Reduction, or the uniqueness of physical reality, remained unexplained however and still stands as one of the greatest challenges to the consistency of physical laws [11].

To begin with, one may recall the main benchmarks of this problem. They began with a model by von Neumann [12], in which the measuring apparatus has only one degree of freedom, such as the position of a pointer along a ruler. The initial state of the pointer is $|a_0\rangle$ at position $a_0$ and the quantum evolution under unitary dynamics yields an outcome

$$|j\rangle|a_0\rangle \to |j\rangle|a_j\rangle, \qquad (1.1)$$

when the initial state $|j\rangle$ that is measured is an eigenvector of the measured observable, and $|a_j\rangle$ is a state of the pointer indicating the corresponding result. If one assumes only two eigenvalues, *i.e.*, $j = 1$ or 2, the model shows that an initial superposition $|\psi\rangle = c_1|1\rangle + c_2|2\rangle$ yields

$$|\psi\rangle|a_0\rangle \to c_1|1\rangle|a_1\rangle + c_2|2\rangle|a_2\rangle. \qquad (1.2)$$

Schrödinger showed the wide generality of superposition after measurement and drew its troublesome consequences, including a striking example where a cat behaves as a pointer [13]. He stressed as a consequence that quantum mechanics does not explain the uniqueness of data and, when uniqueness is imposed through an external reduction of the wave function, the linearity of quantum dynamics is broken.

Wigner extended the analysis to a more realistic case where the state of the apparatus is not pure [14]. When expressed in terms of density matrices, Eq.(1.2) becomes essentially

$$|\psi\rangle\langle\psi|\rho_0 \to |c_1|^2\rho_1 + |c_2|^2\rho_2 + c_1c_2{}^*\rho_{12} + c_1{}^*c_2\rho_{21} \ , \qquad (1.3)$$

and there is no change in the conclusion. Later on, decoherence theory showed that partial traces of the non-diagonal quantities $\rho_{12}$ and $\rho_{21}$ over the environment (*i.e.*, essentially over all the variables apart from the pointer position), vanish rapidly. [1-6], but that does not yield reduction. Allahverdian *et al* reached the same conclusion with no reference to decoherence, in a model showing irreversibility [15].

A common feature of these works (except for the last one) is their exceedingly simple description of a measuring apparatus, which is reduced to the position of an abstract "pointer". Such plainness is particularly puzzling when compared to the sophistication of real measuring apparatuses, so cleverly devised that the invention of some of them was a landmark in the advance of physics. This opposition makes more impressive a work by Bassi and Ghirardi, in which the focusing on pointer is shown sufficient for disproving consistency [16]. Their work involves many refinements and has in principle a wide generality but, .for the present purpose, one will consider it only within the simple framework of Eqs.(1.1-2) as an introduction to the problem of consistency. This framework is standard: An apparatus $A$ measures a system $m$; according to quantum mechanics, a definite Hilbert space $E$ is associated with the system $A + m$, whose dynamics is given by a unitary operator $U(t)$; the position of the pointer is associated with an observable $X$. One assumes that the transitions (1.1) for two eigenvectors $|j\rangle$ result from the evolution of $A + m$ under $U(t)$.

When the Bassi-Ghirardi analysis is reduced to its mathematical content, its conclusion can be stated as an explicit theorem, as follows: If the observable $X$ belongs to a complete set of commuting observables in $E$, there can be no consistency of reduction with the principles of quantum mechanics. More precisely, if $U(t)$ generates a final state $|j\rangle|a_j\rangle$ (with pointer position $x_j$) when acting on an initial state $|j\rangle|a_0\rangle$, then, when the measured initial state $|\psi\rangle$ is a true superposition, $U(t)$ cannot yield a final state in which the position of the pointer is *unique* and close either to $x_1$ or $x_3$.

Bassi and Ghirardi did not consider as an assumption the condition demanding that $X$ belongs to a complete set of commuting observables in $E$, and they considered it as granted. They also did not make explicit the underlying Hilbert space $E$ and the dynamical operator $U(t)$. Nevertheless, their consideration of the case of a non-isolated apparatus in a surrounding universe can make sense only if the overall space $E$ is simply the standard Hilbert space of elementary atoms and particles; whereas $U(t) = \exp(-iHt)$ is associated with the corresponding standard Hamiltonian $H$ (Note: This notation $(E, H)$ relying directly on constitutive particles will always be used in the present paper).

It turns out however that this choice of $(E, H)$ is inconsistent with the existence of a complete set of commuting observables; including $X$, when the pointer is not an abstract concept but a real object belonging to an organized physical apparatus. This prohibition is well known in mathematics [17], and will be explained in Section 2 but, in physical applications, the key word is organization, strongly related to self-organization. These two notions will be explained in more detail later but, for the time being, one may think simply of every solid piece in a clock as something self-organized and of the assembly of all these pieces as the specific organization of the object "Clock". Similar considerations apply to one's preferred detector, but the existence of a loophole in Bassi-Ghirardi's assumptions can be seen more easily in the example of a clock rather than considering immediately a measuring device.

The position of one hand of the clock can stand for an observable $X$ and the question one asks is whether $X$ can be embedded in a complete set of commuting observables in $E$. Since $E$ represents the atoms in the clock and any possible state of these atoms, one can consider a state in which the atoms are organized to build up a very different object, for instance a pendulum. But then, if the position $X$ of a clock hand could belong to a complete sense of commuting observables in $E$, it would make sense for all the state vectors in $E$, including the pendulum. This is obviously absurd. This intuitive argument must be confirmed by a mathematical one, as done in Section 2 where one shows that it is closely related with a mathematical limitation on Dirac's association of classical canonical variables with canonical operators in Hilbert space.

From the standpoint of quantum principles, this restriction has far-reaching consequences, particularly for a basic principle that will be called Axiom A in the present work. It states that a physical system can always be associated with a definite Hilbert space, say $E'$, and when the system is isolated, its evolution is governed by a definite Hamiltonian, say $H'$. The notion of "system" is not however exactly defined in textbooks and a question comes out when one asks its meaning: How can one relate the couple $(E', H')$ to the basic couple $(E, H)$? This question is answered in Section 2 in the case of an organized object, such as a clock or a measuring apparatus before measurement. One finds that self-organization and organization define a definite projection operator $F$ acting on the particle Hilbert space $E$, so that

$$E' = FE, \qquad H' = FHF. \tag{1.4}$$

Since the position *X* of a clock hand (or a pointer) is a collective observable corresponding to a classical observable (in the sense of the correspondence principle or of microlocal analysis [17]), it makes sense in *E'* where it can be included in a complete set of commuting observables. But it is meaningless in *E*, or at least in most part of *E*.

This means that there was a subtle mathematical loophole in Bassi-Ghirardi's argument, or it means at least a significant restriction in the generality of their conclusions. On the other hand, this restriction is also very suggestive. There are experiments in which the system *A+m* is faithfully described by the fundamental couple (*E, H*): they are mostly experiments in quantum optics where *A* and *m* are simple enough to allow a complete mathematical description [18]. They can show decoherence [5], but they never show reduction, as expected from Bassi-Ghirardi's theorem. On the other hand, an organized laboratory apparatus always shows reduction. This means that one can interpret the loophole in Bassi-Ghirardi's argument as yielding a significant suggestion: it indicates a possibly a fundamental role of organization in reduction. This suggestion confirms a previous one by Laughlin [19], who stressed with different arguments a possible role of self-organization.

The present work is therefore mainly an investigation of Axiom A with special attention to measurements. The central problem is the relation between a physical object and its mathematical representation by a couple (*E', H'*). One will insist on a safe definition of a "system", as such, by means of a minimal Hilbert space *E'* accounting for the possible states of the object and a minimal *H'* accounting for its dynamics. In that sense, when a clock is either working, or at a standstill, or out of order, it is represented by three different systems. This sophistication may look pedantic however and is certainly not of much consequence in practice (if it had been, it would have been noticed long ago). It works however as a significant tool for dealing with the problem of consistency, or reduction, and its interest is perhaps restricted to that field, but it is the field in which one is interested here.

One will not pay much attention to generality or completeness in this work, and rather elementary quantum mechanics will be used. This is because the problem of reduction (or the uniqueness of physical reality) is universal, and perhaps not dependent upon new physics. Uniqueness takes place practically under our eyes, at the frontier between the macroscopic domain and the microscopic one, where the basic laws of physics are supposed well known. Moreover, a discovery of a loophole in a no-go theorem does not need much sophistication since pointing it out in a simple case is sufficient. When the question of consistency is reopened on such an occasion, there are two alternatives: Either there is no consistency, after all, and a a better proof of inconsistency must be found, or there is consistency and one must find which mechanism stands behind. The methods in these two attempts have much in common: They cannot rely simply on the consideration of a pointer, but must account for the organization of a real apparatus, and that does not depend whether one expects some final answer or the opposite one.

The content of this paper can then be described as follows: Section 2 deals with the mathematical expression of organization. Section 3 considers then a measurement, with special attention to Axiom A. Before measurement, an apparatus *A* is organized and described by a specific couple (*E', H'*). When it interacts with a measured system *m* in state $|1\rangle$, one must take into account the necessary instability of a reactive part *R* in *A* (for instance the dielectric in a wire chamber when *m* is a charged particle). A new signal (for instance an ionized track) grows exponentially until it can be described by a collective variable $Q_1$ (a localized ionization). When $Q_1$ is large enough, it brings a new contribution to organization so that $A + m$ becomes a new system with a specific couple $(E_1, H_1)$. An analogous phenomenon occurs when $m$ is in state $|2\rangle$: one gets another signal $Q_2$, another system $(E_2, H_2)$ and one finds that $E_2 \neq E_1$, and $H_2 \neq H_1$. When $m$ is in a superposed state $|\psi\rangle$, one must deal with still another system $(E_3, H_3)$, although the same for every $|\psi\rangle$. An essential feature of a measuring apparatus is however that two different Hilbert spaces $E_k$ and $E_l$ (where $k$ and $l$ can take the values 1, 2, or 3) are orthogonal but they are also intersecting (*i.e.* $F_k F_l \neq 0$ for the corresponding projections). This is due to the presence of common self-organized components in the three formal systems, especially the pointer.

One can then investigate the matrix elements of the basic Hamiltonian $H$ between these spaces, including their intersecting and non-intersecting parts. One of them turns out to be especially interesting, because it acts between the two superposed states of the pointer and can lead in principle to fluctuations in the probabilities $p_1$ and $p_2$ of the two channels, strongly suggesting a possible reduction mechanism.

This effect, which is closely related to Laughlin's suggestion, is analyzed in Section 4 in the framework of a simple though standard model of solid-state physics. The conclusions are mitigated: According to a theorem by Pearle [20], these fluctuations could very well yield a complete reduction, with a random final outcome agreeing with Born's probability rule. The fluctuations occur however only when the two states of the pointer, which start from the same initial positions, are still overlapping. This transient situation, which will be called the "proximity period", is rather short and quantitative evaluations are nontrivial. Reduction cannot be excluded nonetheless and this is enough to imply a loophole in Wigner's no-go assertion [14].

If one assumes the validity of consistency, one can apply the present approach to the measurement of an EPR pair by two spacelike-separated apparatuses, which has become recently an essential test for any theory of reduction [21]. This is done in Section 5 and the result agrees with observation. Some considerations are also made in that section on the status of probabilities in quantum mechanics if Schrödinger's dynamics is valid universally, as required by consistency.

**2. Mathematical description of organized systems**

One can distinguish several levels in a mathematical description of organization for a macroscopic quantum system $A$. A mechanical clock, for instance, is made of solid components, and a solid is a typical example of a self-organized system [19]. The first level of organization is therefore self-organization, but there is a second, or higher one, which is the overall organization of the components making a clock. There is another distinction between the two levels of quantum physics and classical physics, which are also both relevant in the case of a clock. Classical dynamics shows how collective coordinates, *i.e.* Lagrange coordinates, describe a constrained motion of the components. There is also a third distinction according to the number of classical coordinates, which is finite in the case of a clock, but in principle infinite, or in practice very large, in fluid mechanics or in electrodynamics. Self-organization enters again there, because of its role in condensed phase of matter, including liquids [19]. The variety is so great that one must choose between generality and simplicity, and simplicity is of course better in a first approach. One will therefore presently restrict the discussion of organization to the case of a clock or a similar mechanical system made of solid pieces.

A solid component of the clock is in the solid state, self-organized according to the rules of solid-state physics. A fundamental property of self-organization is the existence of a bounding surface $\sigma$, which has a permanent shape or a slowly changing one (a clock spring for instance) [19]. The existence of these boundaries, which excludes interpenetration between different components, is essential for a description and an understanding of the clock organization. The introduction of Lagrange coordinates, *i.e.* of a minimal set of variables specifying completely the motion, is also a direct consequence of organization.

When approached in this standard manner, organization appears as essentially classical and macroscopic. One might suspect it, therefore, of being more or less a conceptual structure arising in the mind of an observer. But on the contrary, it is an intrinsic property of the state of the system, as can be seen when considering the quantum density matrix $\rho$ of the clock:

The clock is made of atoms and the significance of boundary surfaces means that a valuable representation of $\rho$, at a specific time, must use the position coordinates $x_k$ of the atoms (leaving aside spin or other quantum numbers). The matrix elements of $\rho$ are then $\rho(\{x_k\}, \{x'_j\})$ and they satisfy Bose-Einstein or Fermi-Dirac symmetries for identical atoms. If one selects some variable $x_1$, associated with the position of a definite atom (for instance some Cu atom), one can trace out all the other variables and obtain a one-atom density matrix $\rho_1(x_1, x'_1)$. It shows the distribution of all the Cu atoms everywhere in the system, because of symmetries. Its diagonal elements $\rho_1(x_1, x_1)$ provide a three-dimensional radiography of the clock and similar mathematical operations on every kind of atom provide many views of the clock showing the shape of its components, their bounding surfaces, as well as the crystal lattices with their geometric symmetries, cell size and orientation, or dislocations and many other detailed features such as chemical composition everywhere. When this description is extended from the level of atoms to the level of nuclei and electrons, one gets also views of the covalent binding of atoms, structure of molecules, and so on. One can therefore consider organization as a set of mathematical structural properties belonging to $\rho$, which remain permanent during a time much larger than the rate of change in the wave functions.

Let one then consider the mathematical representation of organization. In quantum physics, mathematical representations rely on Axiom A involving a couple ($E'$, $H'$). The description of an organized system must give also a meaning to the Lagrange coordinates, particularly the position of a clock hand. But if one takes for $E'$ the Hilbert space $E$ of the constitutive atoms, many different organized systems can be built with the atoms belonging to the clock, as explained in Section 1 where a pendulum was taken as an example.

As already mentioned, mathematics says that the position of a clock hand as an observable has no meaning in $E$ [17]. Classically, the introduction of some Lagrange coordinates $q$ for the clock, together with the right number of microscopic coordinates $z$ for the atoms, amounts to a change of variables $\varphi: \{x_j\} \rightarrow \{z, q\}$. For a pendulum made of the same atoms, one has $\varphi': \{x_j\} \rightarrow \{y', q'\}$. According to Dirac's assimilation of a classical canonical transformation with a quantum unitary transformation, the change of coordinates $\varphi$ would be a canonical transformation and have a quantum version, which would be a unitary transformation $\Phi$. Similarly, $\varphi'$ would have a unitary quantum version $\Phi'$. If so, the product $\Phi\Phi'^{-1}$ would be unitary and this could be used to give a meaning to $q$ as an observable $Q$ acting on a state of the pendulum. But this is wrong, because Dirac's assumption is not universal. The association of a change of coordinates with a unitary transformation is not always valid, as known from Egorov's theorem [22, 23, 17]: $\varphi \circ \varphi'^{-1}$ would have to be 'smooth' enough (*i.e.*, satisfy strong explicit bounds on its derivatives). This not so for different organizes systems and the example of the clock and the pendulum shows that $\varphi \circ \varphi'^{-1}$ is certainly much too 'wild' in such a case.

One must therefore construct explicitly a couple (*E'*, *H'*), necessarily different from (*E*, *H*), to define the clock as a physical system. This is done as follows: One considers the boundary $\sigma$ of a solid component as given by an equation $f(x) = 0$, where $x$ denotes a point in space and $f$ a function. Since the surface $\sigma$ of a component is fuzzy at the atomic scale (typically the Angstrom scale), the functions $f$ and the relative location of the boundaries need not be written exactly and they can be expressed in principle by means of a finite number *N'* of parameters, including *N"* classical Lagrange coordinates $q$ for relative motion of the boundaries. If *N* denotes the number of degrees of freedom for all the particles in the clock, one has *N* >> *N'* >> *N"*. This is sufficient, in principle, to define the mathematical framework of organization in the case of a clock *A*, as follows:

One labels the indices $k$ for the position of atoms $x_k$ according to their organization. One can label for instance all the atoms in a first component, after labeling its lattice sites up to a label *K*, then one labels the atoms in a second self-organized component, starting from index *K* + 1, and so on. If there are liquid or gaseous parts in a system, the atoms are labeled arbitrarily but their nature and number remain fixed inside their own boundaries.

Considering then definite shapes $\sigma$ for the boundaries and definite positions (*i. e.*, definite values $q$ for the Lagrange coordinates), one can construct wave functions $\psi_q$ representing *A* in this configuration. To do so, one starts from any wave function $\psi$ where the variable $x_k$ are labeled and each $x_k$ is located in a definite self-organized component of *A* or in a domain with self-organized boundaries. Two such functions, $\psi$ and $\psi'$, are wave functions of their constitutive particles and, as such, they are endowed with a scalar product and a norm in *E*. Their set is therefore a linear subspace $E_q'$ of *E*. The Hilbert space *E'* can then be defined as the linear closure of the set $\{E_q'\}$ for all the values of $q$, i.*e.* the sum

$$E' = \int E_q dq. \qquad (2.1)$$

*E'* is endowed with the scalar product in *E* and with the corresponding norm. Finally, one uses Fermi-Dirac and Bose-Einstein symmetries to restrict the functions in *E'*. Since *E'* is a linear subspace of *E*, there exists a projection operator *F* such that *E'= F E* and one can define the associated Hamiltonian *H'* by *H' = F H F*.

This construction brings up however a question regarding Lagrange coordinates. The classical quantities $q$ were considered as indices in Eq.(2.1), but they must be also associated with operators $Q$ in *E'*. Two simple examples are given by the orientation of a hand in a clock or the position of a sliding pointer in von Neumann's model. Introducing the center-of-mass of a component through $R = (\sum_k m_k x_k)/(\sum_k m_k)$, where the sum is performed on the atoms in this component, one can use in these examples the difference $Q = R_1 - R_2$, where ($R_1$, $R_2$) denote the center-of-mass positions of the axis around which the hand is rotating and of the hand itself in the first case, or of the ruler and the slider in the second case. *Q* is then a component of this difference, in polar coordinates for the first case, or Cartesian coordinates in the second one.

In an organized state of a clock, the values of an observable *Q* are located around a mean value $q_0$ with an uncertainty $\Delta q$, but how can one prove that the relevant values of *q* in (2.1) are similarly restricted? There is no such proof, because it could only result from a rigorous derivation of organization from the first principles. One may remember in this connection that, at least as far as I know, self-organization of a solid has not even been derived from these principles. One will have therefore to take these properties as granted, or as a physically sensible conjecture, similar to the conjecture underlying self-organization in solid-state physics.

There is still another question regarding linearity. When seen in *E*, the evolution of $\rho$ is given by

$$id\rho/dt = [H,\rho], \qquad (2.2)$$

whereas in *E'* , it is given by

$$id\rho/dt = [H',\rho], \qquad (2.3)$$

but the construction of (*E'*, *H'*) relied on the properties of $\rho$ showing organization and Eq.(2.3) looks in some sense nonlinear, since it relies on an a-priori mathematical relation between $\rho(0)$ and (*E'*, *H'*). This is linked to a wider question concerning the origin of organization: Why does $\rho(0)$ show this organization, and how was it generated [24]? This is a key question. It is obviously related to the direction of time, but one will assume that it is not directly related with reduction.

One can then add also a few remarks, which might be of some help for next steps. The first one is concerned with the use of classical considerations: It turns out that the present approach completes a previous derivation of classical dynamics from the quantum laws [10], since the present construction of (*E'*, *H'*) and of Lagrange observables is in agreement with the assumptions on which this derivation relied. The use of classical concepts is therefore justified in the present framework and this is fortunate since every objective property or datum in an experiment is classical.

The second remark is rather a word of caution. Using the same arguments as in the discussion of $\rho_1$, one can expect that every eigenfunction of the density matrix for a clock will show signs of self-organization and organization. However, an arbitrary wave function in *E'* does not show anything of that kind and stands generally as a purely mathematical state, inaccessible to any experimental construction: nothing like a clock.

The last remark leads to a convenient approximate expression for $\rho$. As mentioned in the introduction, if consistency holds, reduction takes place at the frontier between macroscopic and microscopic physics and it would be wrong, therefore, to ignore powerful procedures that have proved useful in this framework. One of them is the use of boundary conditions to account for physical boundaries. The train-of-wheels and hands of a clock have boundaries σ and it is a common practice to restrict the wave functions of atoms in the various pieces $\alpha$ by boundary conditions. When this is done, $\rho$ can be written in a factorized form

$$\rho = \rho_c \otimes \prod_\alpha \rho_\alpha \ . \qquad (2.4)$$

Here, Π stands for a tensor product of the density matrices for different pieces and $\rho_c$ is the collective density matrix occurring in the derivation of classical dynamics from quantum dynamics [10], and its associated Wigner function involves only the Lagrange coordinates and their associated momenta $(q, p)$. The density matrices $\rho_\alpha$ depend also on the corresponding observables Lagrange canonical coordinates, but in practice only on the classical $(q, p)$.

Eq.(2.4) is approximate. There are small corrections arising from phase correlations between the matrix elements of different density matrices $\rho_\alpha$, not covered by boundary conditions, from phonon exchanges between contiguous pieces, and so on. Anyway, the representation (2.4) is very useful for understanding a complex system and can be used for many purposes with due caution. Its interest is particularly obvious in statistical physics when it is applied to different subsystems in an overall system (it yields for instance most easily an explanation for the additivity of entropy [25]).

## 3. Measurement as an enlargement of organization

Considering that the notion of quantum measurement is sufficiently clear without many words, one will define now a *real measurement* as involving a change in organization. This expression does not mean of course that many remarkable experiments in which no such change occurs deal with unreal events18], but that real measurements are events where the uniqueness of reality is generated, whatever the underlying mechanism. As shown in the introduction together with the complements in Section 2, Bassi-Ghirardi's no-go assertion does not apply to a real measurement, whereas it applies when there is no organization [18]. In this latter case, which may be called a "microscopic measurement", the assertion becomes a theorem and implies that there can be no generation of uniqueness of reality arising from quantum effects, in agreement with observation.

Organization, as seen in Section 2, relies on macroscopic properties, which can be properly expressed by the classical version of quantum mechanics [10]. As a matter of fact, many microscopic events occur spontaneously in the bulk of a macroscopic object, where the state of some microscopic element behaves like a state $|j\rangle$ or $|\psi\rangle$ while the state of another microscopic element behaves like a state $|a\rangle$ in the relations (1.1-2). They do not affect organization. In a real measurement, there is a macroscopic signature, for instance a macroscopic displacement of a pointer, and this property must be understood and mathematically expressed in terms of organization.

One will take a special example, which can be easily generalized in many ways: The measuring apparatus $A$ consists simply of three parts, denoted by $P$, $R$ and $S$; $P$ is a solid pointer whose position involves a unique Lagrange observable $X$. $R$ is a reactive region, which one can suppose unorganized (like a gaseous dielectric in a Geiger counter for instance). $S$ is the support, which encloses for instance the dielectric but also, most importantly, which insures some instability of $R$ under the action of the measured system $m$ (producing for instance an electric field in the counter). $S$ is not necessarily stationary however and carries generally its own evolving Lagrange observables. Before the measurement, $A$ is a well-defined organized system, associated with a couple ($E'$, $H'$).

A real measurement must amplify a quantum event to a macroscopic scale in order to produce a significant effect on the pointer position. Taking as an example the case when $m$ is a charged particle entering into $R$, this amplification process is well known: $m$ ionizes the medium along a track, producing primary electrons, which are accelerated by the electric field, these electrons produce more ions and secondary electrons an so on. One can then formalize this special example in the following way:

Let one assume that, under the influence of either $|1\rangle$ or $|2\rangle$, two distinct tracks are produced in $R$ and the two resulting ionized regions near these tracks are distinct. One can then characterize them by the corresponding number of ions $Q_1$ and $Q_2$, which are integers. Initially, when $A$ is in the state $|a_0\rangle$, one has $Q_1 = Q_2 = 0$. When state $|1\rangle$ is measured, one gets $Q_1 \geq 1$ and $Q_2 = 0$. One can then introduce three Hilbert spaces $E_{R0}$, $E_{R1}$ and $E_{R2}$ for $R$ to describe these mutually exclusive situations and they are orthogonal. When $|1\rangle$ is measured, $Q_1$ grows rapidly and becomes classical. It can then act on the pointer and displace it, for instance through a coupling Hamiltonian $Q_1 P$, analogous to the coupling in Von Neumann's model [12] ($P$ denoting in this special occasion the pointer momentum). The pointer, which was kept fixed in the initial organization of $A$, participates then in a new organization to which the new Lagrange observable $Q_1$ belongs. In the sense of Axiom A, one can then say that the measurement of $|1\rangle$ corresponds to a change of organization and, accordingly, to a change of mathematical system describing $A + m$. It goes without saying that the physical object $A$ is always the same but this convention, far from being a convenience, is a mathematical necessity resulting from the abstractness of quantum mechanics and from the limitations of our intuition and our language (There is by the way some analogy between the present constructions and the method of consistent histories, which was also devised to bypass these limitations. But one will not develop this relation).

The Hilbert space of $A$ before measurement can now be written as

$$E' \equiv E_0 = E_S \otimes E_P \otimes E_{R0}, \qquad (3.1)$$

Similarly, one will use

$$E_1 = E_S \otimes E_P \otimes E_{R1}, \quad E_2 = E_S \otimes E_P \otimes E_{R2}, \qquad (3.2)$$

for the Hilbert spaces that are respectively associated with the measurement of $|1\rangle$ or $|2\rangle$.

One may easily recover the usual standpoint according to which $A$ is considered as a unique system during the process. To this end, one takes $E'' = E_1 + E_2$ as the overall Hilbert space, with $E'' = F''E$. Then one has

$$F_1 = \eta(Q_1)\delta(Q_2)F'', \quad F_2 = \eta(Q_2)\delta(Q_1)F'', \quad F_0 = \delta(Q_1)\delta(Q_2)F'', \qquad (3.3)$$

where $\eta(Q)$ selects the domain $Q > 0$ for the values of $Q$ (remember that these values are integers in the present example). A measurement appears then as an enlargement of organization from zero ionization to either one or two states of ionization (when the initial state of $m$ is either some $|j\rangle$ or $|\psi\rangle$)..

One can also write down the Schrödinger-von Neumann equation for the density matrix of the system $A + m$ during the measurement process. Using $|1\rangle$ and $|2\rangle$ as a basis in $E_m$, one can write it conveniently as

$$\rho_{A+m} = \rho_1|1\rangle\langle 1| + \rho_{12}|1\rangle\langle 2| + \rho_{21}|2\rangle\langle 1| + \rho_2|2\rangle\langle 2|. \tag{3.4}$$

where $\rho_1$ and $\rho_2$ are self-adjoint, positive, with respective traces $p_1$ and $p_2$ such that $p_1 + p_2 = 1$, whereas $\rho_{12}$ and $\rho_{21}$ are adjoint of each other. At time zero, just before measurement, when the initial state of $m$ is $|\psi\rangle\langle\psi|$, with $|\psi\rangle = c_1|1\rangle + c_2|2\rangle$, one has

$$p_1(0) = |c_1|^2, \quad p_2(0) = |c_2|^2. \tag{3.5}$$

One can also introduce different elements for the Hamiltonian, namely

$$H_1 = F_1 H F_1, \; H_1 = F_1 H F_1, \; H_{21} = F_2 H F_1, \; H_{12} = F_1 H F_2. \tag{3.6}$$

Because of the entanglement of $|j\rangle$ with $R_j$, the Schrödinger; Von Neumann equation becomes

$$i d\rho_1/dt = [H_1, \rho_1] + H_{12}\rho_{21} - \rho_{12}H_{21}, \tag{3.7a}$$
$$i d\rho_2/dt = [H_2, \rho_2] + H_{21}\rho_{12} - \rho_{21}H_{12}, \tag{3.7b}$$
$$i d\rho_{12}/dt = H_1\rho_{12} - \rho_{12}H_2 + H_{12}\rho_2 - \rho_1 H_{12}, \tag{3.7c}$$

Taking traces, these equations yield

$$d p_1/dt = 2 \, \text{Im} \, \{Tr(H_{12}\rho_{21})\} = - d p_2/dt \tag{3.8}.$$

These variations of probabilities are not remarkable at first sight. They are familiar in the case of reactions, or when two states are linked through a non-diagonal coupling. Here however, they affect the probabilities of measurement channels, which were always found constant under a linear evolution. This special kind of variations, or fluctuations, can occur only when the measuring apparatus is an organized system and, as discussed in the next section, one knows that if fluctuations are strong enough and random, they can even act as a reduction mechanism [20]. The question whether $H_{12}$ vanishes or not stands therefore at the forefront of the consistency problem.

### 4. Can there be reduction?

Is $H_{12} \neq 0$? The most straightforward method to answer this question is certainly to look for a possible reduction mechanism, because such a search could lead to the discovery of new relevant concepts and, on the other hand, its failure could also enforce a no-go assertion. The problem is however still far from an answer and preliminary investigations show it as a brainteaser in the physics of condensed matter, much exceeding this author's competence. The approach that will be described succeeded nonetheless to show that Laughlin's suggestion is the unique possibility for a mechanism of reduction [19]. It also pointed out an apparently new concept of "proximity", which specifies a very special situation when a pointer is beginning to move and this Laughlin mechanism has a unique opportunity for acting. The short period during which proximity holds is moreover the unique step during a measurement when a loophole can slip into the assumptions of Wigner's no go argument [14]. The aim of the present section will be accordingly to explain these various points as matters of principle, with no pretence at solving the basic problem itself.

*Why Laughlin's suggestion is the only possibility for reduction*

There are interactions between various parts of *A*, but many of them yield only macroscopic classical motion and/or rapidly varying phase correlations, with no effect on the channel probabilities $p_1$ and $p_2$. The stable parts *S* stand only as a support in *A*, and they behave in practice as spectators of the measurement. In a reactive part *R*, there can be quantum interactions between the two signals $Q_1$ and $Q_2$ (which consist of different particles). These interactions imply that different microscopic states of *R* are generated when either $|1\rangle$, $|2\rangle$ or $|\psi\rangle$ are measured (Bassi-Ghirardi's formal description was partly devised to account for this difference [16]). On the other hand, no interaction between two signals (as for instance two distant ionized tracks) can affect their respective probabilities $p_1$ and $p_2$. As for the interactions between *R* and the pointer *P*, they are responsible for the classical action of the signals on the pointer but, except for that, they produce only phase correlations as far as can be seen.

One is thus facing a last possibility, namely an interaction between the two superposed states of the pointer (or more generally of some parts of *A* acting as pointers). This is precisely the case that was pointed out by Laughlin, who stressed a specific resistance, or a reaction, of a self-organized system (or subsystem) against superposition [19]. This Laughlin effect appears here as the unique possible mechanism for reduction (or at least for fluctuations in probabilities), and it could be the main consequence of organization.

*Proximity and a model for Laughlin's effect*

One will consider a special example where the pointer is a moving solid. Its position is a Lagrange observable *X* and one considers two different eigenvalues $x_1$ and $x_2$ of *X*, with a difference $\xi = x_2 - x_1$. The Hamiltonian of the pointer at atomic scale can be expressed for a given value of *x* by [26]:

$$\sum_k p_k^2/2m + \sum_{jk} V_{jk}(x_j - x_k). \qquad (4.1)$$

The positions and momenta of the constitutive atoms have been written here ($x_k$, $p_k$), the functions $V_{jk}$ are the potentials acting between them and the summations range from 1 to *N*, the number of atoms in the pointer.

Laughlin's suggestion was that the state of a self-organized system is necessarily unique, but absolute uniqueness cannot be consistent with quantum mechanics. It belongs, at least implicitly, to a "different universe" and not to one obeying strictly quantum laws. If consistency holds in the sense used here, Laughlin's approach can only be understood as meaning a reaction of a self-organized system against an external action tending to put it into a state of superposition, this reaction bringing the state of the system back to uniqueness. But to speak already of uniqueness would be begging the issue of consistency and the only question in which one can be interested to begin with is the discovery of a quantum mechanism underlying a reaction to superposition.

The cause of the effect is already known. It can only be $H_{12} + H_{21}$ whose meaning is obvious: it is the sum of potential interactions between an atom in the pointer location $x_1$ with another atom in a pointer location $x_2$. The mechanism itself is less obvious. A simple idea would be that, at the beginning of the pointer motion when $\xi$ is still smaller than the cell size $a$ of the solid lattice, an atom in a given lattice position (*e. g.* at $x_1$) interacts with its neighbors in the same lattice, but also with the same atoms in the other lattice position. This second interaction (or mixed interaction) can be supposed responsible for transitions where an atom jumps from a lattice to the other one. Some estimates of orders of magnitude show however that this effect would be probably too small for yielding reduction and there are reasons for expecting a stronger or much stronger effect when a cluster involving several or many atoms, bound together, makes a similar jump. The theory is not much different, at least formally, and one will consider the case of such a cluster.

Quantum jumps show of course a few unusual aspects in this unfamiliar framework. As a general rule, a quantum transition can only take place between two states having wave functions with definite phases, and this condition raises the question of phase correlations in the states of atoms in a pointer. One will assume a finite correlation length of distant atoms and denote it by $\lambda$. It cannot certainly be larger than the mean free path $L$ of a phonon. For instance, the ratio $L/a$ between $L$ and the nearest neighbor distance $a$ is less than 10 for NaCl at room temperature [26], but one will not rely on so specific physical assumptions. One will only introduce a scale of correlation among a hierarchy of systems where the smaller system is a cluster $C$ making a jump. .then there is a sphere $D$ with radius $\lambda$, centered on $C$. Finally, there is the rest of the pointer, denoted by $\overline{D}$. One will assume that $\overline{D}$ stands in practice as a spectator of the quantum transition and enters as a factorized subsystem in an equation similar to (2.4) for the state of the pointer. Most states in $D$ act also as spectators, but as vector states, not through their density matrix. When two such states are simply translated of each other by a distance $\xi$, their scalar product is

$$\langle D_1 | D_2 \rangle = \exp(-N'\xi^2/4\Delta^2) , \qquad (4.1)$$

where $N'$ is the number of atoms in $D$ and $\Delta$ the standard deviation of an atom distance to its lattice site.

The standard theory of quantum transitions yields then the following expression for the probability amplitude of a cluster jump from a state $|C\alpha\rangle$ in the lattice at $x_1$ to a state $|C\beta\rangle$ orthogonal to $|C\alpha\rangle$ and belonging to the other lattice at $x_2$:, while most atoms in $D$ are in the same state in both lattices and unaffected by the jump:

$$\langle C\beta | T | C\alpha \rangle \exp(-N'\xi^2/4\Delta^2) \qquad (4.2)$$

The exponential is due to the atoms in $D$ acting as quantum spectators, as before. The transition matrix $T$ has generally a complex expression in terms of basic potentials, but its meaning is clear: it represents the ejection of $C$ from the initial lattice location at $x_1$ under the action of neighboring atoms in $D$ belonging to the lattice at location $x_2$, somewhat analogous to the ejection of a cherry stone under the pressure of fingers.

The occurrence of the exponential is crucial. It means that no quantum transition can occur when $\xi$ is not significantly smaller than $\Delta$. But $\Delta$ is typically of the order of $10^{-10}$ cm [26], and this means *proximity* in the following sense:: A reduction mechanism originating in quantum transitions can only occur when the distance $\xi$ between the two pointer positions is small as compared to $\Delta$.

Coming back now to Wigner's no-go theorem, one notices that he did not envision this kind of effect, because he concentrated attention on the final states of a pointer, when their positions are clearly separated at a macroscopic scale [14]. This is why he could state that there are no transitions between them, and this specific assertion agrees with the exponential in the amplitude (4.2). The fact that this amplitude does not vanish during the proximity period implies however a loophole in Wigner's argument. This difficulty does not mean however that the conclusion of his argument is necessarily wrong (perhaps reduction is impossible as a quantum effect, after all). Anyway, remembering previous consideration about the Bassi-Ghirardi no-go theorem 16], one can say that no rigorous argument forbids the possibility of consistency.

One may go back therefore to the analysis of possible reduction mechanisms: A calculation of the first factor in the expression (4.2) is not trivial and several problems would have to be understood before one could dare proposing an answer to the reduction problem. There are theoretical problems, and quantitative estimates are also nontrivial but the present paper will remain restricted to matters of principle.

*Self-organization and the behavior of probabilities*
Many kinds of clusters in many different places can participate in quantum transitions, so that one may expect fluctuations in the probability of *presence* of a cluster in a definite position $x_1$ or $x_2$ of the pointer. But what can be their consequence on the probabilities $p_1$ and $p_2$ of the two pointer positions? If $\Delta p$ denotes the change in probability of the cluster in position 1, this question asks for the corresponding change $\delta p$ in the probability $p_1$ of the pointer itself. But it makes sense only for a self-organized system, because a non-organized system such as a molecule cannot let the probability of its atoms increase or decrease. This is possible on the contrary for a solid in which these variations behave essentially like defects. One knows also that self-organization can respond to a fluctuation through some sort of avalanche process and thus spread the defect probability more or less equally over the whole system [24]. If this behavior is also true for local fluctuations in probability for a pointer; one may expect that a local change $\Delta p$ is distributed through the whole pointer to yield a global change $\delta p = (n/N) \Delta p$, where $n$ is the number of atoms in the cluster and $N$ the total number of atoms in the pointer. Of course, this is only a guess, which is far from being proved, but it stresses again that self-organization could be the key for reduction.

*From fluctuations to reduction*
The next question is whether quantum fluctuation processes can yield reduction. Fortunately, the answer was given earlier by Pearle [22]. As a matter of fact, he assumed that the fluctuations arise from nonlinear violations of the Schrödinger equation, but his results do not depend on this assumption, which is irrelevant in the present approach. Pearle's conclusions can be summarized as a theorem, which deals with an arbitrary number of measurement channels and relies on the following hypotheses:: (*i*) The probabilities $p_j$ of the various channels evolve randomly through a Brownian process. (*ii*) The correlation functions of the fluctuations $<\delta p_j\, \delta p_k>$ depend only on time and on the $p_j$'s themselves.(*iii*) If some probability happens to vanish during Brownian motion, it remains zero afterwards. The conclusion of the theorem states that, inevitably, some $p_j$ must finally become randomly equal to 1 while the other $p_k$'s ($k \neq j$) vanish. This outcome is of course reduction, but the most impressive conclusion of the theorem lies in the predicted Brownian probability for getting a specific datum *j*: It is equal to the initial value of the quantity $p_j$ at the beginning of the process, *i.e.*, to the quantum probability $|c_j|^2$ according to Born's rule.

One may notice that Brownian reduction is highly contagious. It can begin in some part of an apparatus, be interrupted and start again in the same place or elsewhere from its latest stage with new correlations, and then it will go on until completion. Contagion means moreover that though quantitative orders of magnitude for the duration of reduction are important, reduction will occur anyway if the fluctuations do not vanish, and even if their achievement is only to kill most branches of Everett's multiple universes [27], long after their birth and leaving a unique surviving branch.

Although Pearle's assumptions are demanding, they are quite sensible in the present approach. Assumption (*i*) would result essentially from the smallness of elementary fluctuations, if they have a Gaussian behavior. Assumption (*ii*) is expected for a quantum jump: Except for matrix elements of interactions, which enter only through their average, a jump probability depends only on the channel probabilities, like in chemical reactions. Assumption (*iii*), which was found sometimes critical in Pearle's works, results in the present case from the irreversibility of signals: When a signal in the reactive part *R* disappears, it cannot be regenerated.

As a conclusion, one can say that the whole problem of consistency hinges on Laughlin's effect: how does a self-organized system react to an outside influence tending to put it into a state of superposition? Reduction and consistency would follow almost automatically and rather easily if this question had a relevant answer, perhaps along the lines that were indicated here. Quantitative estimates are still in progress and will be published later on.

**5. Two aspects of consistency**

One can raise many questions about this approach towards consistency, but two of them are particularly worth answers and comments. The first one, which was asked by Bernard d'Espagnat [a], can be expressed as follows: If consistency holds, dynamics relies entirely on Schrödinger's equation, which is deterministic, and the evolution of a wave function or a density matrix is therefore completely determined by initial values. Why, then, can one predict randomness in measurement data, and what is the origin of this randomness?

This question is a reminder of the deep meaning of reduction, or rather of the uniqueness of reality [11, 28], and it seems rather clear that the present theory is not yet at the right level. Entanglement, proximity, fluctuations and Brownian process, as they were used, look more or less like tricks, rather than acting as foundations. They may be presently the only tools at one's disposal, but there should be a higher conception of the generation of reality if consistency is valid. The question suggests then itself such a change of level and a simple answer, which may be proposed as the following conjecture:

The full details of organization were not used in the previous sections, where only gross features (such as *S*, *P* and *R*) were introduced. But organization is much richer and wider, as one can see when looking for instance at a picture from an electronic or ionic microscope. Surely, so minute details are never exactly reproduced, even in the same apparatus *A,* when two supposedly identical measurements are performed. If the density matrix has a deterministic evolution, according to consistency, the randomness of data could then be attributed to these differences in details. It would mean that the evolution operator *U*(*t*) depends sensitively on the minute details of organization: Mathematically, it would be chaotic.

This means that the issue of a measurement is always unpredictable in practice and macroscopic information about the system can only yield probabilistic predictions, necessarily in agreement with Born's rule. Moreover, a non-probabilistic evolution such as in Eq.(1.1) would mean that the evolution under *U*(*t*) has stable attractors when eigenvectors $|j\rangle$ are measured.

This conjecture has interesting consequences for the status of probabilities and with regard to Einstein's famous sentence "God does not play dice" [29]. The notion of intrinsic quantum randomness becomes replaced, in some sense, by a version of the Laplacian concept of probabilities in which randomness is associated with inaccessibility, or with some insuperable ignorance of exact reality. From a philosophic standpoint, this is remarkable, but with little practical consequence: One can still speak of squared amplitudes as probabilities, since Born's rule is valid, and one can think of such a square as the probability of the corresponding event *if it were measured by an organized apparatus.*

A second question (which was raised by Nick. Herbert [a]) points out that a conjecture must agree with known relevant data, which include now the remarkable experiments by Stepanov *et al* [21] They deal with an entangled EPR pair of photons *γ'* and *γ"* [30]. If one denotes the polarization states as horizontal (*H*) or vertical (*V*), the entanglement is given by the state

$$2^{-1/2}(|H'\rangle|V"\rangle - |V'\rangle|H"\rangle). \qquad (5.1)$$

Two measuring apparatuses *A'* and *A"* measure respectively the polarization of *γ'* and *γ"*. The great interest of the experiment is that *A'* and *A"* are spacelike separated, according to special relativity. Nevertheless, the two measurements always show the correlated outcomes *H'* and *V"*, or *V'* and *H"*, and never *H'* and *H"*, nor *V'* and *V"*. The question is then: Since both measurements are local and separated, why do they show these correlations?

The discussion of this question calls attention to the fact that, till now, a measuring apparatus *A* was considered in this paper as practically isolated. But isolation does not mean only that *A* is isolated:and the whole system *A* + *m* must be isolated. In the present case, when the apparatus *A'* measures the photon *γ'*, it is not isolated from *γ"*, and therefore not from *γ"* and *A"*.

Keeping in mind this reminder, one will use again Pearle's formalism, but with more precision in its foundation. In the case of a measurement by a single apparatus *A* and denoting the fluctuations in the channel probabilities during a short time interval δ*t* by δ*p_j*(*t*) (with *j* = 1 or 2, *H* or *V*), one has

$$\sum_j \delta p_j = 0, \qquad (5.2)$$

because of the normalization property $\Sigma_j p_j(t) = 1$. The average values $<\delta p_j>$ are equal to zero and correlations are generally defined by $A_{jk} = \langle \delta p_j \delta p_k \rangle / \delta t$. Eq.(5.2) implies $\sum_j A_{jk} = 0$. According to Pearle's assumptions, the correlation coefficients $A_{jk}$ depend only on time and on the instantaneous values of the coordinates $p_j$, which will be altogether denoted by $\{p\}$ and which depend also on time. Introducing a Brownian probability distribution $P(\{p\}, t)$ for the values of the random quantities $p_j$ at time $t$, one can write down a Fokker-Planck equation

$$\partial P(\{p\},t)/\partial t = \sum_{jk} \partial_j \partial_k \{A_{jk}(\{p\},t) P(\{p\},t)\}. \qquad (5.3)$$

Pearle's theorem is a consequence of this equation and of the boundary conditions expressing his assumptions. The most significant boundary condition is that no channel with probability zero can be created or recreated during the process, and it was previously justified. In the case of measurements of an EPR pair, the channels *H'* and *H"*, or *V'* and *V"*, have vanishing initial probabilities, and they remain therefore zero. There are however different fluctuations $\delta p_j'(t)$ and $\delta p_j''(t)$ in *A'* and *A"* so that $\delta p_j(t) = \delta p_j'(t) + \delta p_j''(t)$, but they are independent so that $<\delta p_j'(t) \cdot \delta p_j''(t)> = 0$. The only change in Eq.(5.3) is a replacement of the previous $A_{jk}$ by $A'_{jk} + A''_{jk}$ with no change in the conclusion of Pearle's theorem. It may be noticed that this result, which agrees with experiment, is based on the statistical independence of $\delta p_j'(t)$ and $\delta p_j''(t)$, which is insured in any case by the lack of correlation between the initial states of *A'* and *A"*. In other words, the experimental result is explained through the EPR correlation of $\gamma$' and $\gamma$" and the lack of correlation of *A'* and *A"*!

This result is not surprising: Quantum mechanics is non-separable, particularly when the two components of an EPR pair are spacelike separated. If consistency holds, it agrees in principle with every character of quantum mechanics, including non-separability, and nothing else has been used in the present measurement theory, except of course for organization.

## Acknowledgements

The long construction of this theory would not have been possible without remarks, criticisms or encouragement by Stephen Adler, Roger Balian, Angelo Bassi, Robert Dautray, Bernard d'Espagnat, Jacques Friedel, Frank Laloë, Philippe Nozières, Philip Pearle and Heinz-Dieter Zeh. I thank Mireille Calvet for her help with the manuscript.

## Notes and references